%% file: conference_101719.tex
\documentclass[conference]{IEEEtran}
\IEEEoverridecommandlockouts
\usepackage{cite}
\usepackage{amsmath,amssymb,amsfonts}
\usepackage{algorithmic}
\usepackage{graphicx}
\usepackage{textcomp}
\usepackage{xcolor}
\usepackage{multirow}
\usepackage{RobStd}
\usepackage{multirow}
\usepackage{booktabs}
\usepackage{enumitem}
\def\BibTeX{{\rm B\kern-.05em{\sc i\kern-.025em b}\kern-.08em
    T\kern-.1667em\lower.7ex\hbox{E}\kern-.125emX}}
    
\IEEEoverridecommandlockouts
\IEEEpubid{\makebox[\columnwidth]{978-1-6654-3922-0/21/\$31.00 $\copyright$2021 IEEE \hfill} \hspace{\columnsep}\makebox[\columnwidth]{ }}

\begin{document}
\title{Application-driven Design Exploration for Dense Ferroelectric Embedded Non-volatile Memories}

\author{

Mohammad Mehdi Sharifi$^{*\infty}$, Lillian Pentecost$^\dagger$$^{\infty}$, Ramin Rajaei$^{*}$, Arman Kazemi$^{*}$, Qiuwen Lou$^{*}$, \\Gu-Yeon Wei$^\dagger$, David Brooks$^\dagger$, Kai Ni$^{\psi}$, X. Sharon Hu$^{*}$, Michael Niemier$^{*}$, Marco Donato$^{\gamma}$\\
\normalsize $^{*}$University of Notre Dame, \normalsize $^\dagger$Harvard University, \normalsize $^\psi$Rochester Institute of Technology, \normalsize $^{\gamma}$Tufts University, $^{\infty}$equal contribution
 }

\maketitle

\begin{abstract}
The memory wall bottleneck is a key challenge across many data-intensive applications. Multi-level FeFET-based embedded non-volatile memories are a promising solution for denser and more energy-efficient on-chip memory. However, reliable multi-level cell storage requires careful optimizations to minimize the design overhead costs. In this work, we investigate the interplay between FeFET device characteristics, programming schemes, and memory array architecture, and explore different design choices to optimize performance, energy, area, and accuracy metrics for critical data-intensive workloads. From our cross-stack design exploration, we find that we can store DNN weights and social network graphs at a density of over $\mathbf{8MB/mm^{2}}$ and sub-2ns read access latency without loss in application accuracy.
\end{abstract}


\section{Introduction}

Data-intensive applications such as deep neural networks (DNNs) and graph analytics have emerged as dominating workloads in the current computing landscape and have influenced many research efforts and advances in computer architecture in the past few years. Specialized hardware architectures deliver outstanding performance and computational efficiency. 
However, the size and complexity of critical workloads continues to outstrip available on-chip memory capacity, making data movement and efficient on-chip storage an outstanding challenge in scaling these applications.

Embedded non-volatile memories (eNVM) offer a promising alternative to traditional on-chip SRAM \cite{ibm_nvm_survey}. 
Smaller memory cell size and the ability to store multiple bits in a single memory cell can translate to a denser array implementation, which can be optimized to store entire DNNs on-chip \cite{maxnvm, memti}. 
In addition to storage density improvements, non-volatility increases energy efficient due to lower leakage power and the ability to retain information in power-off state for intermittent computing. 
The advantages of eNVMs are often countered by lackluster write performance and reduced reliability,
which may be exacerbated with multi-level cell programming.

In today's varied application space, meeting different memory read and write access patterns requires careful consideration of eNVM memory characteristics. 
An application-driven approach can provide a clear path towards design optimization targeting a specific application and coincidentally offer insight regarding which device-level specifications such as reliability, endurance, write performance, and cell area should be further improved to serve a wider set of applications.   


\begin{figure}[t]
    \centerline{\includegraphics[width=0.85\columnwidth]{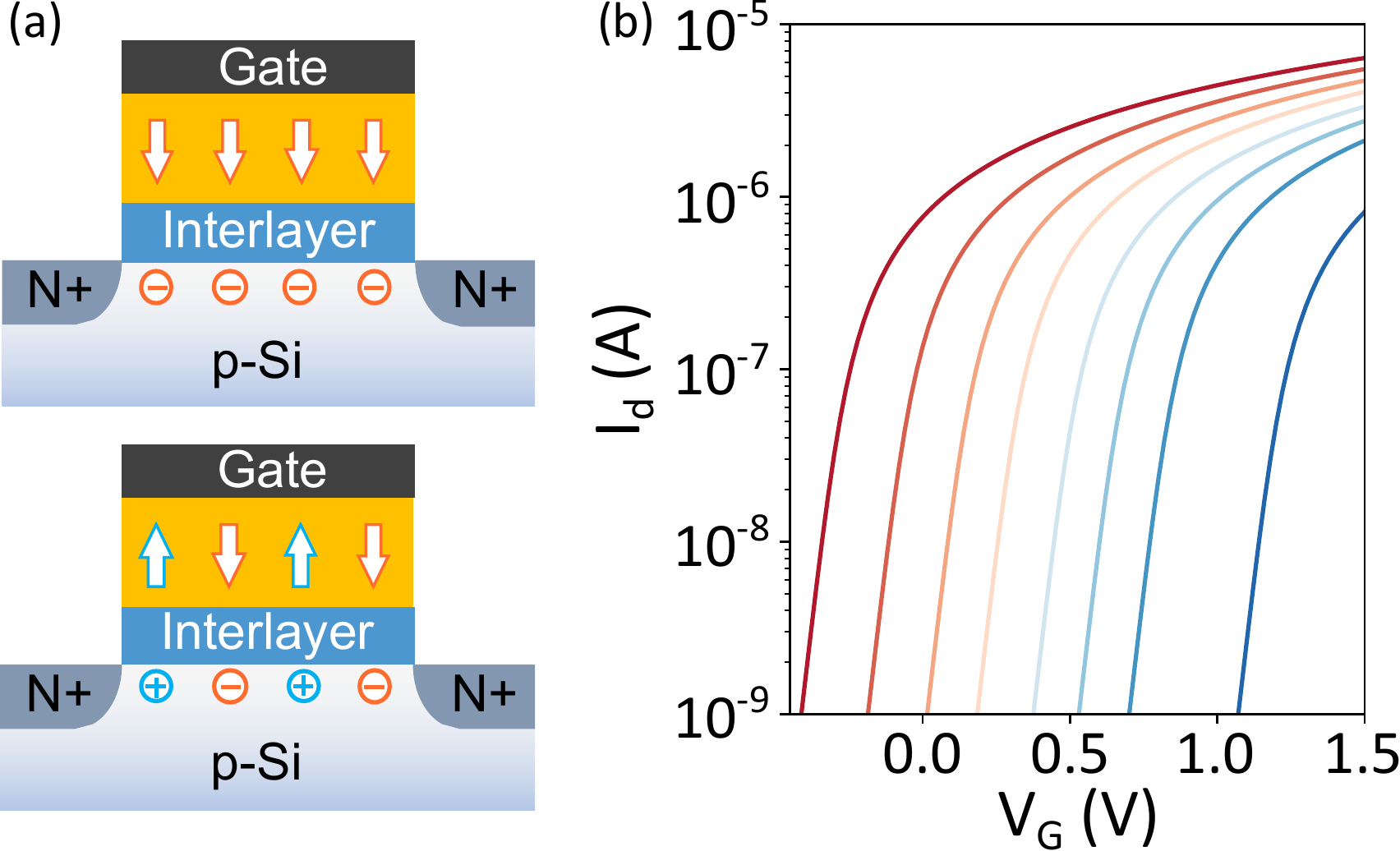}}
    \caption{(a) FeFET device; (b) Transfer characteristics of a FeFET device in 8 different resistance levels for 3-bit-per-cell programming.}
    \label{fig:FeFET_device}
    \vspace{-0.1in}
\end{figure}

In this work, we advance the study of a particularly compelling eNVM proposal, Ferroelectric FET (FeFET), by modeling and quantifying the impact of device-level design choices such as size and programming scheme on the accuracy of target workloads.
FeFET memories promise dense storage with competitive read characteristics
, but their viability and achievable density in a realistic application use case, such as DNN inference or graph analytics, has not yet been explored.

We begin by evaluating the device-to-device (D2D) variation for different FeFET cell areas using model presented in \cite{ni2020FeFETD2D}. 
Based on these preliminary device-level results, we investigate and develop custom programming schemes to increase the programming reliability for single-level cell (SLC) and multi-level cell (MLC) configurations. 
To quantify the relationship between resulting device characteristics and memory array performance, we extend the NVSim tool to support a FeFET cell model \cite{nvsim}. 
Additionally, we develop a FeFET fault model based on the variability both at the memory and sensing circuitry level, and we extend an application-level fault injection framework to evaluate the impact of FeFET device size and programming scheme on application accuracy.

This paper provides the first comprehensive design exploration and evaluation of FeFET-based eNVMs for realistic use-cases such as DNN weight storage and graph analytics.
More specifically, we make the following contributions:
\begin{itemize}[noitemsep,topsep=0pt,parsep=0pt,partopsep=0pt,leftmargin=*]
\item We perform a device-level study, modeling the effects of memory cell size and programming schemes on the fault rate, energy, and write latency for single and multi-level (2-bit and 3-bit) cell implementations; 
\item We evaluate array-level implications of different FeFET cell characteristics by modeling and optimizing memory architecture and sensing circuit to minimize read errors; 
\item We introduce a FeFET fault model and an application-level fault-injection study to quantify the application accuracy across cell-level and memory-level characteristics for a collection of data-intensive workloads.
\end{itemize}

\section{Background}

\subsection{Ferroelectric FET memories} 
\begin{figure}[t]
    \centerline{\includegraphics[width=1\columnwidth]{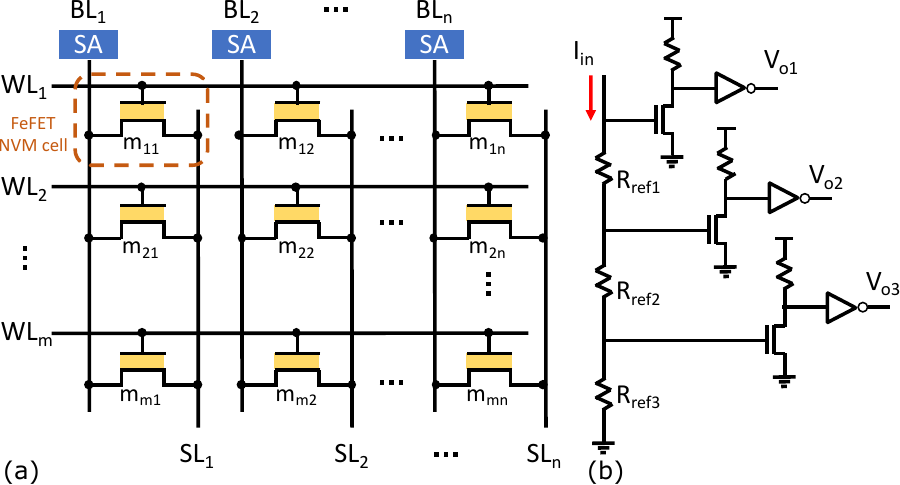}}
    \caption{(a) AND array memory architecture; (b) Customized multi-level sensing circuit for four programmed (2-bit) MLC.}
    \label{fig:AND_Array_SA}
    \vspace{-0.2in}
\end{figure}

The FeFET device (Figure \ref{fig:FeFET_device}(a)) is made by replacing the normal gate dielectric in a MOSFET with a ferroelectric layer \cite{jerry2018ferroelectric}. For this reason, FeFET devices can be easily integrated in existing CMOS processes, especially when high-$k$ dielectrics such as hafnium oxide can be used as the ferroelectric layer~\cite{hafnium}.
The polarization of the ferroelectric layer can be controlled by applying different voltages to the gate of the device, where this polarization sets the threshold voltage ($V_{th}$) of the device. 
Recent work~\cite{reis2019design,ali2019multilevel,dunkel2017fefet} uses FeFETs for ultra-dense, low-leakage, and fast memories.  
This work employs 1FeFET AND arrays (Figure \ref{fig:AND_Array_SA}(a), where the source lines (SLs) connect the elements along the columns), which are more promising than the other types of FeFET memories (e.g., NOR arrays, where the SLs connect the elements along the rows) for two reasons: first, the cell size is equal or even smaller than NOR arrays; second, the parallel bit-lines and source-lines reduce the write disturbance \cite{ni2018write}. Note that the write disturbance can be further mitigated by applying a half-bias scheme (e.g., applying $V_{write}/2$ to deselected cells) \cite{reis2019design}.

\subsection{Multi-level NVM storage} 
Storing multiple bits in a single memory cell is desirable for increasing storage density, and has been demonstrated using many eNVM devices, including resistive random access memories (ReRAM), phase change memories (PCMs), charge-trap transistors (CTT), and FeFETs~\cite{ibm_nvm_survey,ctt}. 
In order to discriminate among different programmed levels, the memory uses analog to digital converters (ADCs). 
The ADCs translate each programmed $I_{D}$ target level to the corresponding binary word~\cite{MLCtradeoffs}. 
One key challenge in MLC storage is 
the higher susceptibility to D2D variations, making each programmed level less reliable \cite{ali2019multilevel}.
In addition, the higher area and power overhead of the ADCs discriminating between more current levels  can limit the relative benefits of MLC implementations compared to single-level-cell (SLC) storage.

\subsection{Target applications} 
\label{subsec:apps}

We evaluate the impact of FeFET reliability characteristics on target applications that are (1) data-intensive in terms of required capacity and read bandwidth and (2) have infrequent or highly batched write accesses.
In embedded and mobile SoCs, the high density and compelling efficiency of FeFET memories could be crucial in enabling on-device or otherwise power-efficient execution of these critical applications, and performance will not be debilitated by potentially long writes.

Deep Neural Networks (DNNs) for vision tasks and natural language processing (NLP) are well-studied driving applications that have demonstrated resilience to MLC eNVM storage \cite{maxnvm}.
DNN inference performance and power-efficiency benefits significantly from increased on-chip memory density, particularly for storage of weight parameters that are infrequently updated.
Thus, we evaluate the viability of FeFET MLC memories for two representative DNN workloads: ResNet18 for image classification (CiFar10), and the ALBERT transformer-based model for natural language understanding (MNLI) \cite{resnet, albert}.
Another critical category of data-intensive workloads are those that operate on large graphs \cite{graphs}, such as search tasks on social network graphs, though resilience of graph formats to MLC eNVM storage has not been addressed in prior work.
Thus, we additionally evaluate the viability of FeFET MLC memories for storing social network graphs with sample graphs representing Wikipedia article voting patterns and anonymized social circles from Facebook \cite{snapnets}.

\begin{figure}[t]
    \centerline{\includegraphics[width=1\columnwidth]{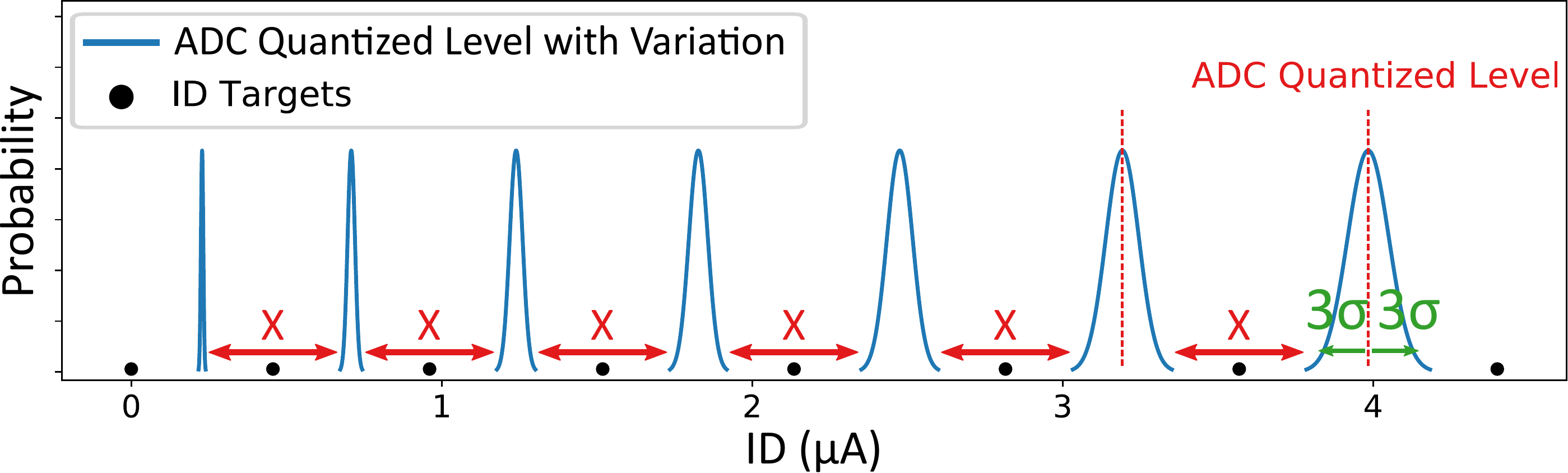}}
    \caption{ADC quantized levels with variation as a Gaussian function with 3$\sigma$ deviation of 5\% and target currents for 3-bit FeFET MLC.}
    \label{fig:ID_targets}
    \vspace{-0.2in}
\end{figure}

\section{Methodology}

\subsection{FeFET multi-level current model} 
\label{sec:MonteCarloModel}


FeFETs, like other emerging devices, suffer from D2D variation, an issue that severely limits their scalability both in terms of physical dimension and multi-level programming. 
In our study, we account for D2D variations by employing a stochastic model of polarization switching in FeFET devices~\cite{ni2020FeFETD2D}. The model uses Monte Carlo sampling on independent polarization domains in the ferroelectric layer. 
In doing so, it captures the essential behaviors required to perform scalability and reliability studies on FeFET memories, including (i) D2D variation as the cell size changes (different number of $10nm\times10nm$ domains); (ii) stochasticity of domain switching; and (iii) the accumulation of domain switching probability when a train of pulses are applied to the gate of the FeFET device. Also, it is worth noting than this model has been experimentally validated.
Using this model, we can effectively project the impact of write schemes and device size on D2D variation, and, consequently, the impact of D2D variation on application level accuracy.

\subsection{Memory array architecture model} 
We extend NVSim~\cite{nvsim} to extrapolate our device-level characterization study to memory array architecture. 
NVSim does not natively support FeFET memory cells, so we add a customized memory cell definition. 
Moreover, we modify NVSim to estimate the costs of MLC sensing circuitry. 
These additions are discussed in the following subsections.  
\subsubsection{NVSim MLC FeFET model}  
    
We integrate a new memory cell definition using the energy, delay, and area results collected from SPICE simulations. 
Our evaluation considers two programming schemes, which are discussed in detail in Section~\ref{subsec:schemes}.
The write energy and latency for the write-verify scheme use the average number of set and reset programming pulses computed by sampling 1500 FeFET cells for D2D variation using the model discussed in~\ref{sec:MonteCarloModel}, in addition to estimated energy and latency due to write circuitry.

\begin{figure}[t]
    \centerline{\includegraphics[width=0.75\columnwidth]{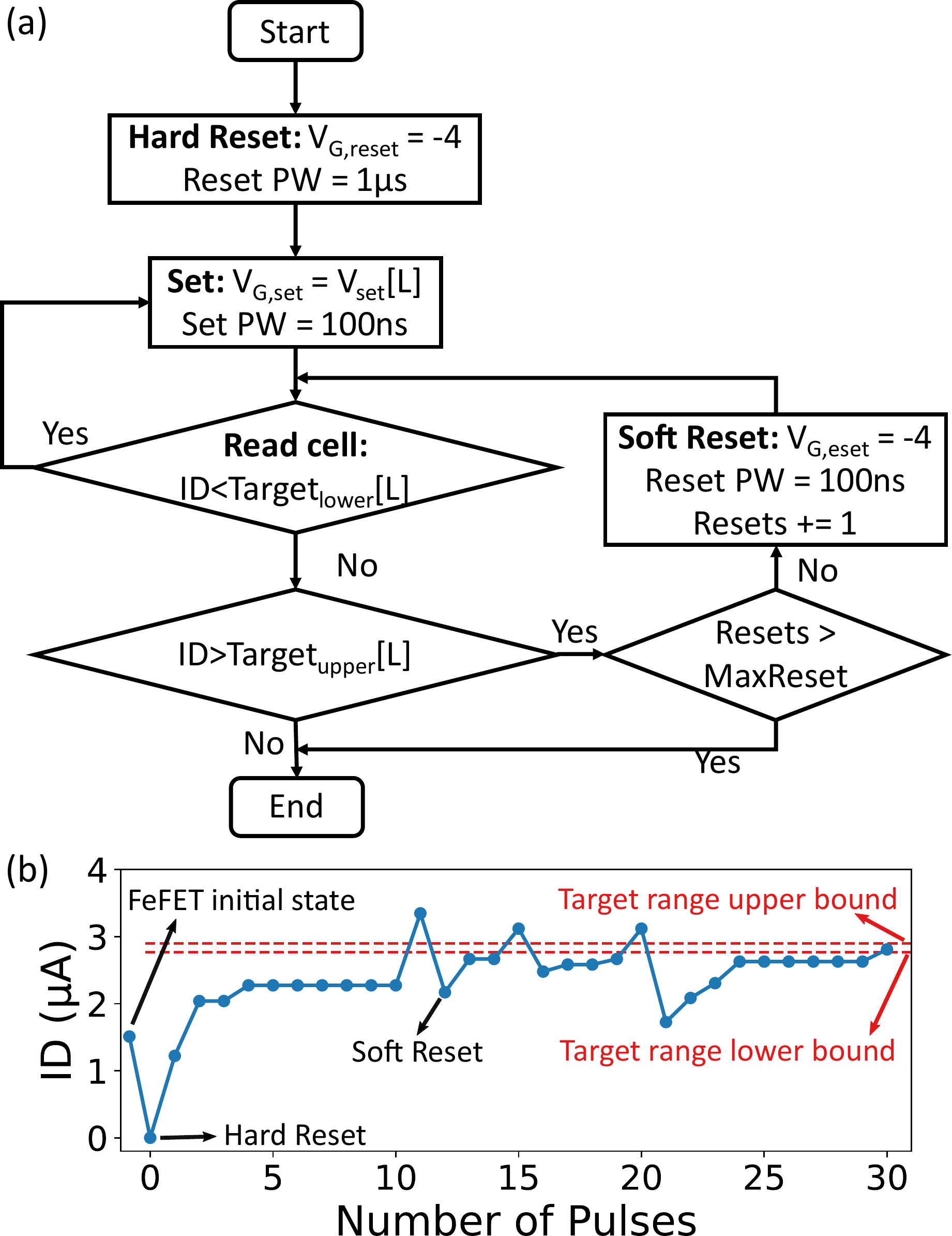}}
    \caption{(a) Proposed write-verify programming scheme; (b) Example of pulse by pulse level tuning process of an FeFET device into the target range.}
    \label{fig:scheme}
    \vspace{-0.2in}
\end{figure}


An additional change to improve model fidelity is that FeFETs fabrication requires adding a ferroelectric layer on top of a traditional MOSFET gate, so the gate equivalent oxide thickness is increased and, therefore, the gate capacitance. We captured this in NVSim by setting FeFET gate capacitance 1.73$\times$ bigger than the default CMOS gate capacitance.

\subsubsection{Sensing circuit design} 
For a 1-bit read operation, we employ a simple voltage-based sense amplifier, which we characterize in SPICE. 
The same sense amplifier model is employed to build a parallel sensing scheme for MLC operation. 
In a parallel sensing scheme, the current from a memory cell is compared against $2^n-1$ reference levels and returned as an $n$-bit binary word.
Figure~\ref{fig:AND_Array_SA} depicts the circuit schematic of the sensing circuit for a 2-bit read operation.
The operation is similar to a Flash ADC and is similarly affected by D2D variation, which in turn has an impact on memory reliability. 
We model the effects of D2D variations on transistor dimensions, resistance values, etc., as a Gaussian function with a 3$\sigma$ deviation of 5\% \cite{pang2009measurement,razi2020design}. As a result, the quantized levels show variability proportional to the threshold currents. 
Based on this constraint, we propose spacing the MLC programming currents such that the sensing threshold distributions are equally spaced (Figure~\ref{fig:ID_targets}). 
This approach uses the extra margins in low-current levels to distribute the read errors due to sensing threshold shifts more evenly across the entire programming window. 
The energy, latency, and area for the resulting design is incorporated in NVSim to model sensing at the memory array architecture level.


\label{sec:sensAmp}

\begin{figure}[t]
    \centerline{\includegraphics[width=0.8\columnwidth]{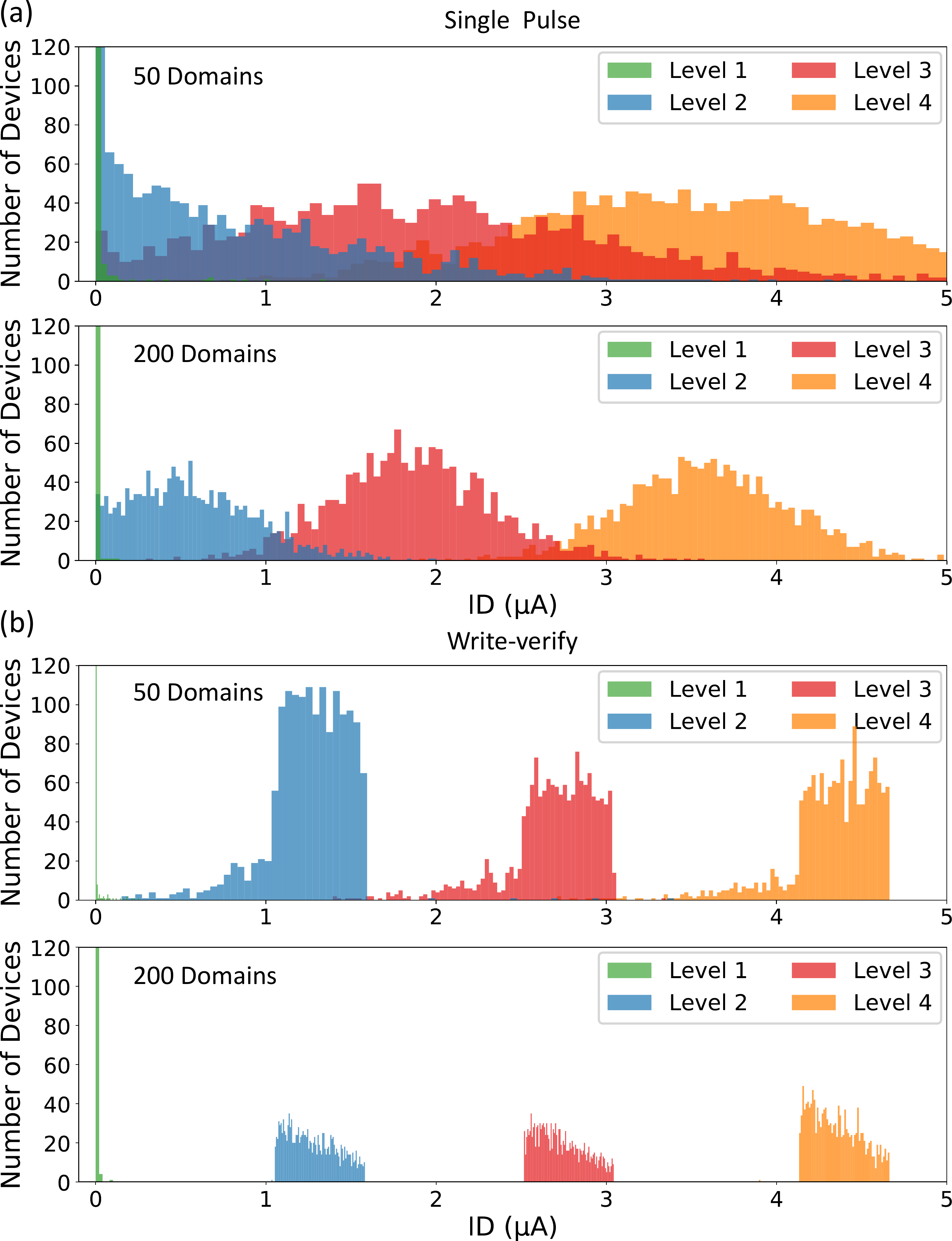}}
    \caption{(a) Current distribution of 2-bit FeFET MLCs with single pulse programming for 1500 cells; (b) Current distribution of 2-bit FeFET MLCs with write-verify programming for 1500 cells.}
    \label{fig:Dist_plot}
    \vspace{-0.2in}
\end{figure}

\subsection{Cross-application fault injection framework} 


Evaluating device non-idealities at the application level requires a balance of fault model accuracy and simulation performance. 
As the focus of this work is on DNN and graph applications, we leverage two popular Python packages, namely PyTorch~\cite{pytorch} and SNAPpy~\cite{snap} to execute representative workloads.
Previous work has explored the evaluation of faults in DNN models~\cite{ares} and extended the study to eNVM errors in DNN applications~\cite{ctt-dac}. We extend existing MLC eNVM fault modeling to simulate FeFET memories based on the model discussed in Section~\ref{sec:MonteCarloModel} and ADC quantized level variations. 
Moreover, we use a general input interface to process parameters from both DNN and graph workloads. 
For each target application, the fault injection tool reads in the application data to be stored in FeFET-based memory and applies a quantization transform followed by MLC encoding based on the selected configuration. 

Stored data values are assigned to FeFET current levels, and we sample the current based on the FeFET Monte Carlo model. The currents computed in this first sampling process are then compared against current sensing thresholds which are also sampled based on ADC variations. The sensed currents are then converted back to quantized values and used to repeat workload execution and evaluate the effect of memory faults on the target application.

\begin{figure}[t]
    \centerline{\includegraphics[width=0.9\columnwidth]{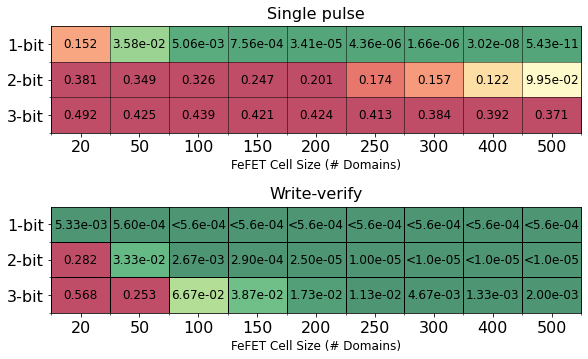}}
    \caption{Shmoo plot of maximum read fault probability as a function of cell size and bits per cell for single pulse vs. write-verify programming.}
    \vspace{-1em}
    \label{fig:shmoo_mlc}
\end{figure}

\section{FeFET optimization study}
\subsection{Programming schemes} 
\label{subsec:schemes}
We consider two approaches for programming multi-level FeFETs, (i) single pulse; and (ii) write-verify. 
The single-pulse approach first performs a hard reset of the device (a -4$V$, $1 \mu s$ pulse), followed by one pulse of amplitude dictated by the target level. 
While appealing in terms of simplicity, the single-pulse scheme is not robust to D2D variation for smaller FeFET cells, as evident in the overlap of resulting current distributions in Figure \ref{fig:Dist_plot}(a) for 50-domain cells.

The write-verify scheme proposed in this work (Figure~\ref{fig:scheme}(a)) helps combat D2D variations by producing tighter level distributions. After a hard reset, cells are supplied with shorter $100 ns$ fixed-amplitude pulses. 
After each pulse, the device current is measured against a reference to determine whether the target current has been exceeded, and a soft reset pulse of the same duration is applied to correct any overshoot. 
As shown in Figure \ref{fig:scheme}(b), a soft reset pulse does not return the device to its initial state because of the shorter applied pulse. 
This gives us more control to set the device to a target range. 
A similar strategy has been applied to RRAM devices \cite{shim2020two}, but, to the best of our knowledge, the impact of this approach has not been studied for FeFETs. 
Additionally, our use of fixed-amplitude pulses reduces the overhead of the write circuitry. 
Conversely, other emerging devices such as ReRAM and PCM require variable amplitude and width in programming pulses to reach the desired current targets \cite{shim2020two, chen2020parallel}.

Our evaluation shows that only a small subset of devices (less than 0.1\% for 200-domain cells) fails to reach the target range after 10 soft reset cycles. Therefore, we fix the maximum number of soft resets to prevent prohibitively long and energy-consuming programming sequences. 
The programming sequence is terminated either if the measured device current is within a target range, or if the number of applied soft reset pulses reaches its maximum count (Figure \ref{fig:scheme}(b)). 


Figure \ref{fig:Dist_plot} shows the current distribution of 1500 2-bit FeFET MLCs programmed using single-pulse and write-verify programming schemes. Figure \ref{fig:Dist_plot}(a) shows that there are large overlapping areas between different levels if the single-pulse scheme is used. The level distribution overlap worsens when the number of domains is reduced from 200 to 50. Figure \ref{fig:Dist_plot}(b) shows that using write-verify scheme, overlaps can be mitigated for 200 domain cells. However, even with the write-verify scheme, there is some overlap in 50 domain cells.


\begin{figure}[t!]
\centering
\includegraphics[width=0.48\textwidth]{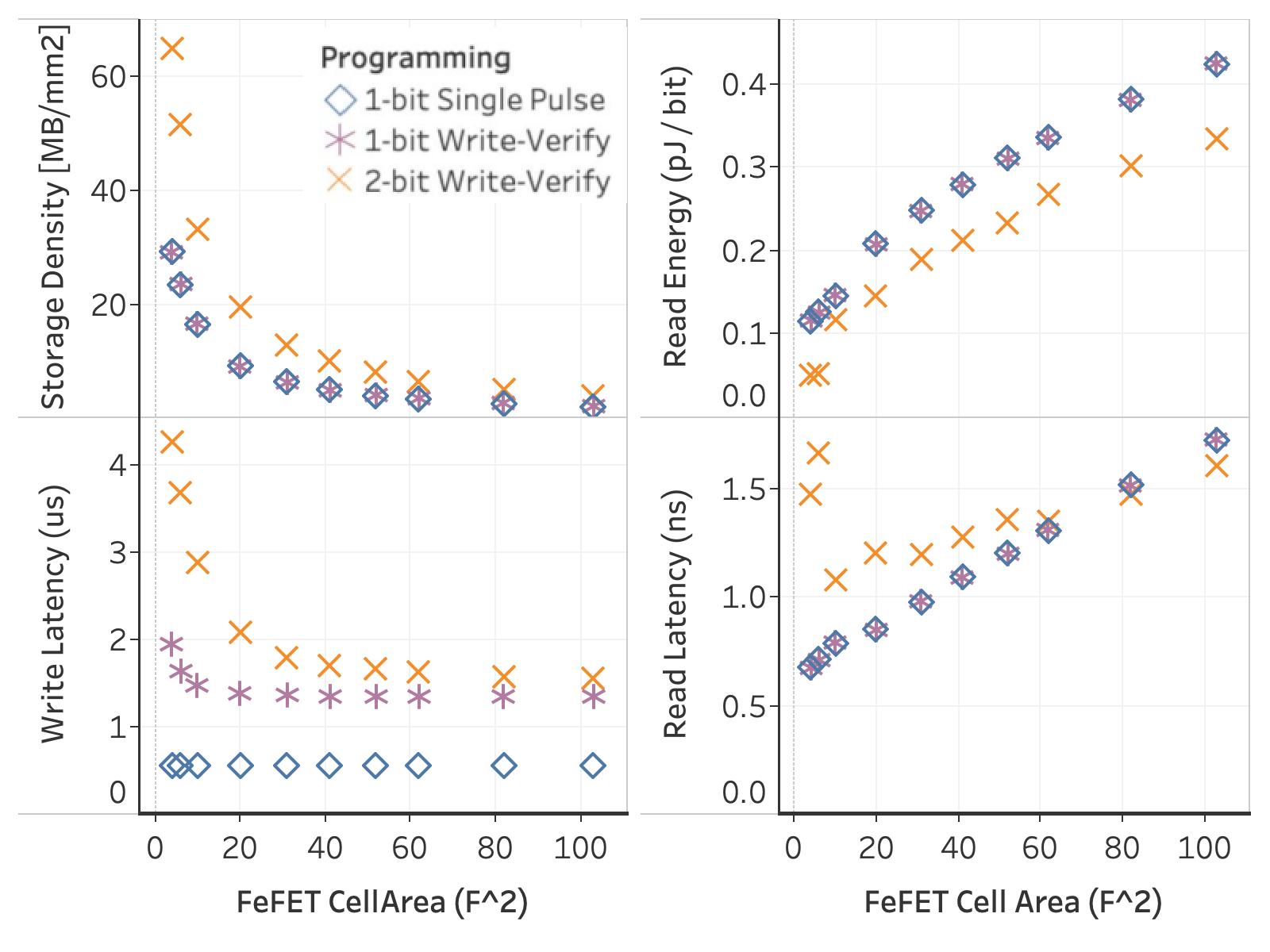}
\vspace{-10pt}
\caption{For a 4MB capacity array with fixed optimization target (read energy-delay product) and word width (64b), we observe varying storage density and performance for different cell sizes and programming choices.}
\label{fig:array}
\vspace{-0.2in}
\end{figure}

\subsection{Memory design trends}
For each of the programming schemes discussed in Section \ref{subsec:schemes}, we carefully consider the fault characteristics of the FeFET memory as we vary the number of programmed levels per cell and the cell size.
The highest resulting inter-level fault rate per cell design and programming scheme is shown in Figure \ref{fig:shmoo_mlc}.
Our fault model additionally captures the asymmetry of inter-level fault rates (i.e., when it is more likely for the programmed level to be mis-read as a lower level than a higher one), which is increasingly apparent in 2-bit and 3-bit programming.
Thus, the raw maximal fault rate alone does not capture the complexity of the potential impact of MLC FeFET behavior on application accuracy.

The resulting memory array storage density and performance characteristics for both programming schemes and across cell sizes is shown in Figure \ref{fig:array}.
We note that for a fixed capacity and NVSim optimization target, 2-bit storage provides strictly better density and read energy independent of programming choices.
Though write-verify and MLC programming increase array write latency (Figure \ref{fig:array}, bottom left), we highlight that the compelling read characteristics and impressive storage density inspire us to investigate the viability of these memories for critical, data-intensive workloads.

As discussed in Section \ref{subsec:apps}, there are multiple classes of important workloads that can be tolerant of long write latency and would benefit immensely from increased storage density and efficiency.
However, these workloads may exhibit different resilience to MLC fault characteristics.
In light of the relatively high error rates for MLC programming across cell size shown in Figure \ref{fig:shmoo_mlc}, it is critical that we evaluate application-level implications to identify the densest allowable storage scheme without degrading application accuracy.



\begin{figure}[t!]
\centering
\includegraphics[width=0.48\textwidth]{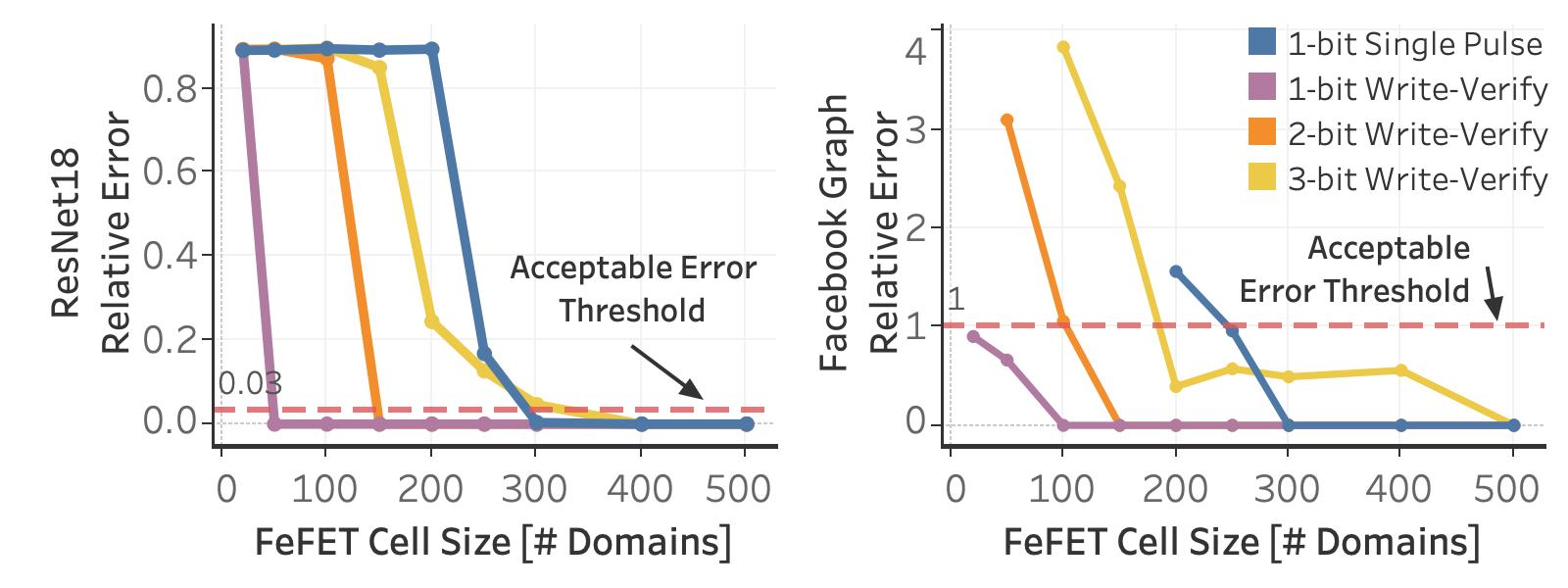}
\vspace{-10pt}
\caption{Fault injection studies are performed per cell size and programming scheme to evaluate resulting application error, as measured by relative degradation from baseline accuracy.  The minimum cell size that keeps relative error below the acceptable threshold dictates densest storage per scheme.}
\label{fig:app_accuracy}
\vspace{-0.2in}
\end{figure}

\section{Application-level evaluation}
\label{sec:eval}


\subsection{DNN Weight Storage} 

We evaluate two DNN workloads which differ both in terms of target application and eNVM utilization. 
The first is image classification using ResNet18. 
In this case, we consider a system in which the full model is stored in the FeFET memory. Figure \ref{fig:app_accuracy}, left, shows how application error relative to baseline classification accuracy increases significantly as cell size decreases across programming schemes.  For 1-bit, single-pulse programming (no verify), the minimum FeFET cell size that preserves inference accuracy is 300 domains. 
Introducing the write-verify scheme allows the memory cell size to be scaled more aggressively, with  1- and 2-bit storage displaying minimum cell sizes of 50 and 150 domains, respectively. 

The second DNN we consider is natural language inference running on ALBERT. 
BERT-based DNNs are often re-used across multiple tasks (e.g., translation vs. sentiment analysis), so we propose to store the shared embedding parameters in FeFET memory, while the task-specific portion of the network is stored separately, an approach that has proven successful in maximizing energy efficiency for multi-task DNN inference~\cite{memti}. In this case, the 1-bit, single-pulse configuration requires 200 domains, while for the write-verify cases, 1-, 2-, and 3-bit configurations can be scaled down to 50, 150, and 150, respectively. Compared to the relative vulnerability of 3-bit MLC for ResNet18, partitioning ALBERT to store more vulnerable parameters in SRAM provides the opportunity for more aggressive scaling in terms of FeFET cell size.    

\input{tables/app_results_v2}

\input{tables/array_results}

\subsection{Social Network Graph Storage} 
To determine the viability of different FeFET cell designs and programming schemes for graph analytics tasks, we measure the average impact to the accuracy of a set of breadth-first search queries on the example graphs as a proxy for maintaining network structure and resilience to faults across many integral graph processing kernels \cite{graphs}.
We store input graphs described in Section \ref{subsec:apps} as adjacency matrices and perform fault injection to quantify resilience to MLC storage.  The relative error increase for the Facebook graph, for example, (Figure \ref{fig:app_accuracy}, right), demonstrates workload-specific sensitivity to FeFET fault characteristics distinct from ResNet18 in terms of allowable cell size.
The minimum cell size per workload and per programming scheme is summarized in Table \ref{tab:apps}.

Due to the varying sparsity and structure of the two graph workloads studied, we observe different fault resilience in the 3-bit configuration, and the Wikipedia voting graph degrades in terms of average query accuracy when using a cell with fewer than 400 domains.
Each mis-read value in the adjacency matrix corresponds to an erasure or erroneous addition of a connection between two graph nodes, and it is possible that each mis-read value has a disproportionate impact on a graph that is less clustered and more sparsely affiliated (as in the Wikipedia graph) than one with more disparate, strongly connected social circles (as in the Facebook graph), but further analysis is required to justify this argument.



\subsection{Application-Provisioned FeFET Arrays}

\begin{figure}[t!]
\centering
\includegraphics[width=0.95\columnwidth]{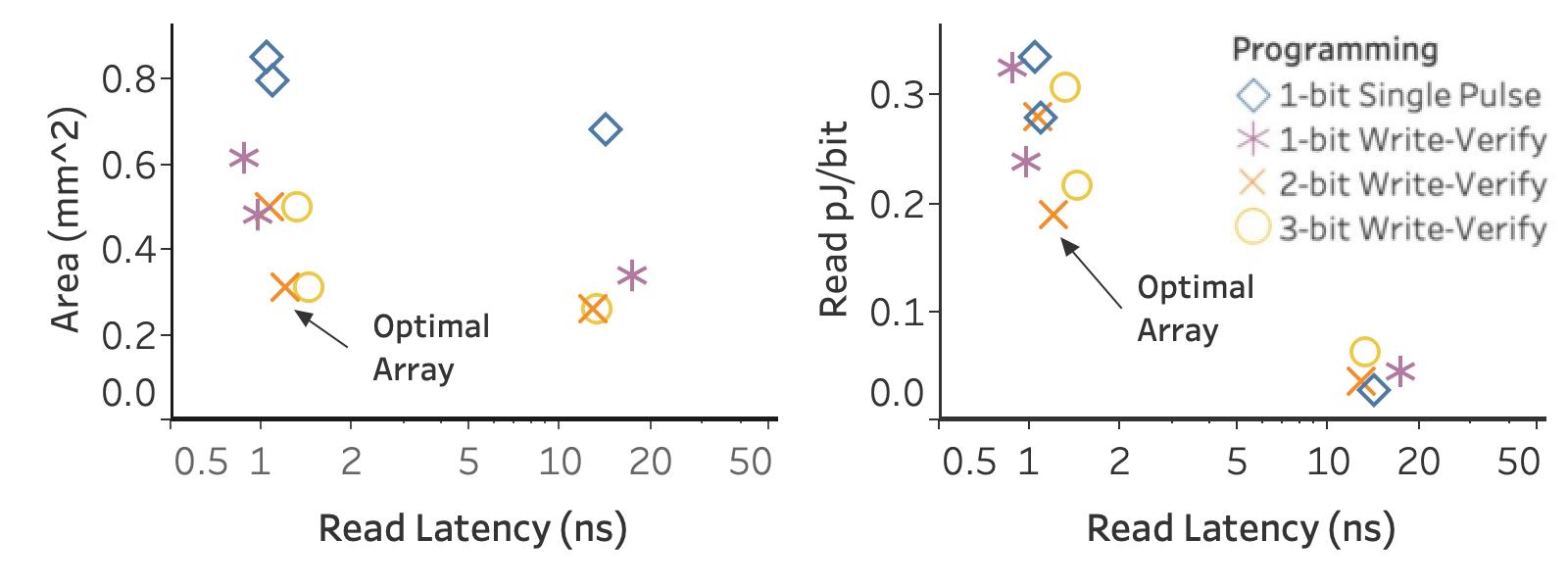}
\vspace{-10pt}
\caption{For an example workload (ALBERT, 4MB capacity), we show key characteristics for FeFET memory arrays with acceptable application accuracy across programming schemes and NVSim optimization targets.
From these, we select optimal array configurations summarized in Table \ref{tab:summary} that minimize array area, read latency, and read energy per access.}
\label{fig:albert}
\vspace{-0.2in}
\end{figure}

Though all of the examined workloads are amenable to MLC FeFET storage, in order to derive the most efficient FeFET storage solution per application, we must consider the achievable zero-accuracy-loss storage density in conjunction with memory array performance, area, and energy characterization.
For example, we find that SLC FeFET under the write-verify scheme can be safely used down to a cell size of 50 domains or smaller (Table \ref{tab:apps}), so this programming scheme consistently provides highly dense, zero-accuracy-loss memory.
This is highlighted for an example application (ALBERT) in Figure \ref{fig:albert}.

Figure \ref{fig:albert} also demonstrates that 2-bit storage is a compelling design in terms of density and read characteristics, with 3-bit showing competitive performance in all metrics, though not providing the pareto-optimal configuration highlighted in Figure \ref{fig:albert} and summarized for each workload in Table \ref{tab:summary}. 
The storage density achieved here is a full order of magnitude higher than that of 16nm SRAM, in which a 4MB array requires about 4$mm^2$ according to NVSim evaluation, cited as a reference point in Table \ref{tab:summary}.
Similarly, the observed read latency and energy per bit tends to match or improve upon SRAM.
We repeat analysis of area, latency, and read energy per bit of provisioned memory arrays per-application to identify FeFET memory arrays that minimize latency and energy.
In this way, the FeFET arrays highlighted in Table \ref{tab:summary} will cause minimal impact to workload execution time while providing high density, energy-efficient embedded storage.

It is important to note that these design points are not fully optimized, which indicates that even higher density could be achieved with architectural enhancements \cite{maxnvm}. 
While these optimizations are outside of the scope for this paper, we want to highlight two possible additions.  
First, incorporating lightweight error mitigation or error correction strategies could enable use of even higher density storage solutions, such as 3-bit MLC with smaller cell sizes.
Previous work considering fault-prone MLC eNVM storage unveiled that hand-tuned design studies can maximize total storage efficiency.  Second, more sophisticated programming schemes (e.g., increasing pulse amplitudes  \cite{shim2020two}) could be used to further tighten the current distribution of MLC FeFET configurations. However, more complicated programming schemes would add area, latency, and energy overhead due to write circuitry. 

\section{Conclusion}
This work is a demonstration of the viability of MLC-programmed FeFET eNVMs with an application-driven evaluation of these memories in light of DNN and social network graph resilience and memory access patterns. 
Our methodology and presented studies provide exposure to the critical trade-offs among write costs, storage density, fault characteristics, and application impact that will drive future development of highly-dense FeFET memory solutions. 
In our simulations, we expose the fault characteristics and performance implications of different programming schemes for a set of interesting data-intensive workloads. Our results show that MLC FeFET storage can be effectively leveraged for both deep neural network inference and graph processing tasks, achieving even in the absence of more involved optimizations.

\section*{Acknowledgment}
This work was supported by ASCENT and ADA research centers, JUMP centers cosponsored by SRC and DARPA. 
This work is also sponsored in part by E2CDA-NRI, a funded center of NRI, a SRC program sponsored by NERC and NIST. 

{\footnotesize
\bibliography{bibliography}}
\bibliographystyle{./my_bib_style.bst}

\end{document}

%% file: tables/app_results_v2.tex
\begin{table}[b]
\vspace{-0.2in}
\caption{Summary of minimal cell size in number of domains per programming scheme and bits per cell (BPC) without application accuracy degradation, as discussed in Section \ref{sec:eval}.}
\centering
\begin{tabular}{|c|c|c|c|c|c|}
\hline
\textbf{BPC} &  \textbf{Programming}  &  \textbf{ResNet18} &  \textbf{ALBERT} &  \textbf{Wiki} &  \textbf{Facebook} \\   \hline \hline
1 & Single Pulse      &  300  &  200  &  250  & 250 \\ \hline
1 & With Verify       &  50   &  50   &  50   & 20  \\ \hline
2 & With Verify       &  150  &  150  &  150  & 150 \\ \hline
3 & With Verify       &  400  &  150  &  400  & 200 \\ \hline
\end{tabular}
\label{tab:apps}
\end{table}

%% file: tables/array_results.tex
\begin{table*}[t!]
\centering
\caption{Summary of best programming schemes and full memory array characteristics provisioned per workload.}
\begin{tabular}{|c|c|c|c|c|c|c|c|c|}
\hline
\textbf{Workload} & \textbf{Optimal Scheme} & \textbf{\begin{tabular}[c]{@{}c@{}}Storage \\ Requirements\end{tabular}} & \textbf{\begin{tabular}[c]{@{}c@{}}\# Cell \\ Domains\end{tabular}} & \textbf{\begin{tabular}[c]{@{}c@{}}Array Area \\ ($mm^2$)\end{tabular}} & \textbf{\begin{tabular}[c]{@{}c@{}}Read Latency \\ ($ns$)\end{tabular}} & \textbf{\begin{tabular}[c]{@{}c@{}}Read Energy \\ ($pJ$ / $bit$)\end{tabular}} & \textbf{\begin{tabular}[c]{@{}c@{}}SET Latency \\ ($\mu s$)\end{tabular}} & \textbf{\begin{tabular}[c]{@{}c@{}}SET Energy \\ ($pJ$ / $bit$)\end{tabular}} \\ \hline \hline
ResNet18          & SLC with Verify         & 24MB                                                                     & 50                                                                  & 1.686                                                               & 1.866                                                             & 0.461                                                              & 1.47                                                              & 0.745                                                                 \\ \hline
ALBERT            & MLC2 with Verify        & 4MB                                                                      & 150                                                                 & 0.313                                                            & 1.20                                                       & 0.189                                                                  & 1.80                                                          & 0.438                                                            \\ \hline
Wikipedia         & MLC2 with Verify        & 6MB                                                                      & 150                                                                 & 0.682                                                             & 1.268                                                               & 0.278                                                                    & 1.80                                                        & 0.422                                                             \\ \hline
Facebook          & MLC2 with Verify        & 2MB                                                                      & 150                                                                 & 0.241                                                              & 0.976                                                           & 0.169                                                               & 1.80                                                                & 0.234                                                                \\ \hline \hline
SRAM (16nm)          & Reference Point       & 4MB                                                                      & N/A                                                                 & 3.9                                                              & 1.3                                                           & 0.5                                                              & 0.001                                                               & 0.5                                                                  \\ \hline
\end{tabular}
\label{tab:summary}
\vspace{-0.1in}
\end{table*}